\documentclass{svjour2}                    
\smartqed  
\usepackage{graphicx}
\newcommand{\tr}{\mathop{\rm tr}}
\newcommand{\Tr}{\mathop{\rm Tr}}

\newcommand{\dif}{\mbox{\rm d}}
\newcommand{\ci}{\mathop{\textrm{i}}}

\journalname{General Relativity and Gravitation}
\begin{document}

\title{Rainich theory for type D aligned Einstein-Maxwell solutions
}

\titlerunning{Type D aligned Einstein-Maxwell solutions}        

\author{Joan Josep Ferrando \and \\ Juan Antonio S\'aez }

\authorrunning{J.J. Ferrando \and J.A. S\'aez} 

\offprints{Joan J. Ferrando}          
\institute{J.J. Ferrando \at Departament d'Astronomia i
Astrof\'{\i}sica,
\\Universitat
de Val\`encia, E-46100 Burjassot, Val\`encia, Spain.\\
\email{joan.ferrando@uv.es}   \and J.A. S\'aez \at
Departament de Matem\`atiques per a l'Economia i l'Empresa,\\
Universitat de Val\`encia, E-46071 Val\`encia, Spain.\\
\email{juan.a.saez@uv.es}}
\date{Received: date / Revised version: date}
%
\maketitle

\begin{abstract}
The original Rainich theory for the non-null Einstein-Maxwell
solutions consists of a set of algebraic conditions and the Rainich
(differential) equation. We show here that the subclass of type D
aligned solutions can be characterized just by algebraic
restrictions.
\end{abstract}

\section{Introduction}
\label{intro}

The energy tensor $T$ associated with an electromagnetic field $F$
solution of the source-free Maxwell equations ({\it Maxwell field}),
$\nabla \cdot F = 0, \ dF=0$, is divergence-free, $\nabla \cdot T =
0$. Conversely, if $T$ is a conserved symmetric tensor, what
additional conditions must it satisfy in order to be the energy
tensor of a Maxwell field? This problem was posed and solved by
Rainich \cite{rai} for non-null fields obtaining, as a consequence,
a fully geometric characterization of the non-null Einstein-Maxwell
solutions. It is worth pointing out that the Rainich approach
\cite{rai} also includes other interesting results about the
principal planes of a non-null Maxwell field (see, for example
\cite{fsY} for a detailed analysis).

Here we are interested in the following points of the Rainich work:
(i) to express Maxwell equations for the Maxwell-Minkowski energy
tensor, such, to obtain the algebraic conditions and the additional
differential restrictions for a conserved symmetric tensor to be the
energy tensor of a Maxwell field, and (ii) to write all these
conditions, via Einstein equations, for the Ricci tensor considered
as a metric concomitant.

The Rainich answer to the the first point can be summarized in the
following \cite{rai}:

\begin{theorem}  \label{teo-rainich-1}
{\bf (Rainich, 1925)} A symmetric tensor $T$ is the energy tensor of
a non-null Maxwell field if, and only if, it satisfies the algebraic
conditions:
\begin{equation} \label{Rainich-1-a}
\tr T = 0, \qquad \quad   4 T^2 = \tr T^2\, g \not= 0, \qquad \quad
T(x,x) > 0   \label{alg}
\end{equation}
and the differential ones:
\begin{equation} \label{Rainich-1-b}
\nabla \cdot T = 0, \qquad \quad   \mbox{\rm d} \Psi[T] =0
\end{equation}
where $x$ is an arbitrary time-like vector and the Rainich 1--form
$\Psi[T]$ is given by
\begin{equation}  \label{Rainich-1-c}
\Psi[T] \equiv \frac{6}{\tr T^2} *(\nabla T \cdot T) \label{PsiT}
\end{equation}
\end{theorem}
The operators and products used in this theorem and hereafter are
explicitly defined in Appendix A.

The answer to the second point of the Rainich work follows easily
from the above theorem. Indeed, under the traceless condition,
Einstein-Maxwell equations state that the Ricci tensor and the
energy-momentum tensor differ in a constant. Thus, algebraic
conditions (\ref{Rainich-1-a}) apply on the Ricci tensor. Moreover,
field equations imply the conservation condition and only the
Rainich (differential) equation must be imposed. Consequently, we
have the following \cite{rai}:

\begin{theorem}  \label{teo-rainich-2}
{\bf (Rainich, 1925)} A metric tensor $g$ is a non-null
Einstein-Maxwell solution if, and only if, its Ricci tensor $R
\equiv R(g)$ satisfies the algebraic conditions:
\begin{equation} \label{Rainich-2-a}
\tr R = 0, \qquad \quad   4 R^2 = \tr R^2\, g \not= 0, \qquad \quad
R(x,x) > 0   \label{alg}
\end{equation}
and the differential one:
\begin{equation} \label{Rainich-2-bc}
\mbox{\rm d} \Psi(R) =0 \, , \qquad  \Psi(R) \equiv \frac{6}{\tr
R^2}
*(\nabla R \cdot R)
\end{equation}
where $x$ is an arbitrary time-like vector.
\end{theorem}

Rainich-like characterizations have been considered for other matter
models. Thus the Mariot-Robinson theorem \cite{mariot,robinson} is
the starting point in characterizing the null Einstein-Maxwell
solutions. Several subsequent works have contributed to this goal
(see, for example, Ref. \cite{bartrum,ludwig}). In a different
framework, in considering a hydrodynamical matter model, the local
thermal equilibrium condition has been expressed in terms of energy
tensor variables and a fully geometrical description of the
thermodynamic perfect fluid solutions has been obtained \cite{cf-1}.

It is worth remarking that in a Rainich-like approach we can limit
the set of metrics where it applies or we can consider matter models
with additional physical properties. Thus, the Rainich theory for
the ideal gases in local thermal equilibrium has been useful in
characterizing the ideal gas Stephani universes \cite{cf-2}. The
conceptual and practical interest of these `reduced' Rainich
theories motivates the goal of this work: {\it the characterization
of the type D aligned Einstein-Maxwell solutions}.

This family of metrics contains significant solutions like the
Reissner-Nordstr\"om and the Kerr-Newman black holes and their
vacuum limit, the Schwarzschild and Kerr solutions, as well as,
other well-known space-times that generalize them, such as the
charged Kerr-NUT solutions.

In posing this Rainich problem we look for the Einstein-Maxwell
solutions in the set of the Petrov-Bel type D metrics whose
principal 2-planes are those of the (non-null) Maxwell field. At
first glance, we could solve this question by adding to the
(algebraic and differential) Rainich conditions (\ref{Rainich-2-a})
and (\ref{Rainich-2-bc}), the complementary algebraic conditions
that state:

(i) The Weyl tensor is of Petrov-Bel type D.

(ii) The Weyl and Ricci tensors have aligned principal planes.

In this work we find explicit expressions in terms of the Ricci and
Weyl tensors for the conditions (i) and (ii). But our results go
quite a lot further. Indeed, we show that, under the alignment
restriction (ii), the algebraic Rainich conditions
(\ref{Rainich-2-a}) imply the type D requisite (i). On the other
hand we show that, by adding a simple algebraic constraint, the
algebraic conditions imply the (differential) Rainich equation
(\ref{Rainich-2-bc}). This means that no differential conditions are
necessary to characterize the type D aligned Einstein-Maxwell
solutions, that is, the two differential conditions
(\ref{Rainich-1-b}), the divergence-free equation and the Rainich
equation, follow from the field equations. More precisely, in this
work we show the following:

\begin{theorem}  \label{teo-our-1}
A metric tensor $g$ is a type D aligned Einstein-Maxwell solution
if, and only if, its Ricci and Weyl tensors,  $R \equiv R(g)$ and $W
= W(g)$ satisfy the {\em algebraic} conditions:
\begin{equation} \label{Rainich-3-a}
\tr R = 0, \qquad \quad   4 R^2 = \tr R^2\, g \not= 0, \qquad \quad
R(x,x) > 0   \label{alg}
\end{equation}
\begin{equation} \label{aligned-3}
R_{(\alpha}^{\mu} P_{\beta) \mu \gamma \delta} = 0 \, , \qquad A^2 +
B^2 \not= 0 \, , \qquad B^2 + ( 3 A - \tr R^2)^2 \not= 0,
\end{equation}
where
$$
\begin{array}{l}
\displaystyle P \equiv W - \alpha G - \beta \eta \, , \qquad \quad
\alpha \equiv - \frac{AC+BD}{A^2+B^2} \, ,  \qquad \beta \equiv
\frac{AD-BC}{A^2+B^2} \, ,\\[3mm]
A \equiv \frac12 \Tr W^2 , \ \ B \equiv \frac12 \Tr (W \circ
*W),  \ \ C \equiv \frac12 \Tr W^3,  \ \ D \equiv \frac12 \Tr (W^2 \circ
*W) ,
\end{array}
$$
and where $G = \frac12 g \wedge g$ is the metric on the {\rm
2}--forms space, $\eta$ is the metric volume element and $x$ is an
arbitrary time-like vector.
\end{theorem}
The operators and products used in this theorem and hereafter are
explicitly defined in Appendix A.

This work is organized as follows. In section
\ref{sec-algebraic-conditions} we give an alignment condition
between the Weyl and Ricci tensors and show that, under this
restriction, the algebraic Rainich conditions (\ref{Rainich-2-a})
imply that the Weyl tensor is Petrov-Bel type D. In section
\ref{sec-differential-conditions} we analyze the differential
Rainich equation (\ref{Rainich-2-bc}) and we show that it is
identically satisfied by adding an algebraic scalar constraint to
the algebraic restrictions studied in section
\ref{sec-algebraic-conditions}. Moreover we justify the above
theorem that characterizes the metric tensors $g$ and we give the
procedure to obtain the Maxwell field $F$ which complete the
Einstein-Maxwell solution. Section \ref{sec-ending} is devoted to
considering the counterpart solutions with cosmological constant and
to analyze the vacuum limit. Finally, in order to clarify the
notation used in the work, the operators and products are explicitly
defined in Appendix A.

\section{Algebraic conditions}
\label{sec-algebraic-conditions}

Let $(V_4,g)$ be an oriented space-time of signature $\{ -, +,+,+
\}$, and let $R$ and $W$ be the Ricci and Weyl tensors, defined from
the curvature tensor as given in \cite{kra}.

A non-null electromagnetic field $F$ takes the canonical expression
$\, F = e^{\phi}[\cos \psi \, U \! + \sin \psi \,*U]$, where $U$ is
a simple and unitary 2--form that we name {\it geometry} of $F$,
$\phi$ is the {\it energetic index} and $\psi$ is the {\it Rainich
index}. The intrinsic geometry $U$ determines a 2+2 almost-product
structure defined by the time-like plane $V$ generated by the null
eigenvectors of $F$ and whose volume element is $U$, and its
space-like orthogonal complement $H$. They are the {\em principal
planes} of $F$ (or of $U$). If $v$ and $h= g-v$ are the respective
projectors and $\Pi = v-h$ is the {\it structure tensor}, we have $v
= U^2$ and $h = -(*U)^2 $, where $*$ denotes the Hodge dual operator
and $U^2 = U \cdot U$ (see Appendix A).

The ({\it Maxwell-Minkowski}) energy tensor $T$ associated with an
electromagnetic field $F$ is minus the traceless part of its square
and, in the non-null case, it depends on the intrinsic variables
$(U, \phi)$:
\begin{equation}
T \equiv -\frac{1}{2}[F^2+*F^2] = -\frac{1}{2}e^{2\phi}[U^2+*U^2] =
-\kappa \Pi   \label{TF}
\end{equation}

The symmetric tensor (\ref{TF}) has the principal planes of the
electromagnetic field as eigenplanes and their associated
eigenvalues are $\pm \kappa$, with $2 \kappa = \sqrt{\tr T^2} =
e^{2\phi}$. Then, $T$ satisfies the conditions (\ref{Rainich-1-a}).
Conversely, if a symmetric tensor satisfies the first two algebraic
restrictions in (\ref{Rainich-1-a}), then it is, up to sign, the
traceless part of the square of the simple 2--form $F^{\circ}$ given
by:
\begin{equation} \label{F0}
F^{\circ} = F^{\circ}(T) \equiv  \frac{Q(X)}{\sqrt{2Q(X,X)}} \, ,
\qquad  Q \equiv T \wedge g -\frac{1}{\sqrt{\tr T^2}} (T \wedge g)^2
\,   \label{alg2}
\end{equation}
where $X$ is an arbitrary 2--form, $\wedge$ denotes the double-forms
exterior product, and $S^2$ means the square of a double 2--form $S$
considered as an endomorphism on the 2--form space (see Appendix A).

But, in order to guarantee the physical meaning of an energy tensor
$T$ we must also impose the energy conditions on it. Under the first
two algebraic restrictions in (\ref{Rainich-1-a}) the Pleba\'nski
energy conditions reduce to the third one. Consequently,
(\ref{Rainich-1-a}) gives the {\it algebraic} characterization of a
Maxwell-Minkowski energy tensor. Then, taking into account that the
energy tensor determines the electromagnetic field up to duality
rotation, we can state:

\begin{lemma} \label{lemma-FT}
A symmetric tensor $T$ is an energy tensor of non-null
electromagnetic type if, and only if, it satisfies the algebraic
conditions {\em (\ref{Rainich-1-a})}.

If $T$ satisfies {\em (\ref{Rainich-1-a})} we can obtain the simple
{\rm 2}-form $F^{\circ}$ given in {\rm (\ref{alg2})}. Then, for any
scalar $\psi$, the electromagnetic field $F = \cos \psi F^{\circ} +
\sin \psi
*F^{\circ}$ has $T$ as its energy tensor.
\end{lemma}

In order to analyze the conditions for a space-time to be a
Petrov-Bel type D solution we now introduce a necessary notation. A
self--dual 2--form is a complex 2--form ${\cal F}$ such that $*{\cal
F}= \textrm{i}{\cal F}$. We can associate biunivocally with every
real 2--form $F$ the self-dual {2--form ${\cal
F}=\frac{1}{\sqrt{2}}(F-\textrm{i}*F)$. We here refer to a
self--dual 2--form as a {\it SD bivector}. The endowed metric on the
3-dimensional complex space of the SD bivectors is ${\cal
G}=\frac{1}{2}(G-\textrm{i} \; \eta)$, $\eta$ being the metric
volume element of the space-time and $G$ the metric on the space of
2--forms, $G=\frac{1}{2} g \wedge g$.

Every double-2--form, and in particular the Weyl tensor $W$, can be
considered as an endomorphism on the space of the 2--forms. The
restriction of the Weyl tensor on the SD bivectors space is the {\em
self-dual Weyl tensor} and is given by:
$$
{\cal W} \equiv {\cal G} \circ W \circ {\cal G} = \frac12(W - \ci
*W)
$$

The algebraic classification of the Weyl tensor $W$ can be obtained
by studying the traceless linear map defined by the self--dual Weyl
tensor ${\cal W}$ on the SD bivectors space \cite{petrov-W,bel,fms}.
The characteristic equation reads $\, x^{3}-\frac{1}{2} a x
-\frac{1}{3} b =0  \, $, where the complex invariants $a$ and $b$
are given by:
\begin{equation} \label{invariants}
a \equiv  \Tr {\cal W}^2  \, , \qquad b \equiv \Tr {\cal W}^3 .
\end{equation}

In a Petrov-Bel type D space-time the self-dual Weyl tensor has a
double eigenvalue and a minimal polynomial of degree two, and it
admits the canonical expression \cite{fms}:
\begin{equation}  \label{type-D-canonica}
{\cal W} = 3 \rho \, {\cal U} \otimes {\cal U} + \rho \, {\cal G} \,
, \qquad \rho \equiv - \frac{b}{a}
\end{equation}
where ${\cal U}$ is the normalized eigenbivector associated with the
simple eigenvalue $-2\rho$. The {\em canonical bivector} ${\cal U}$
determines the 2+2 structure $\Pi = 2 {\cal U} \cdot \bar{\cal U}$,
where $\bar{\ }$ denotes the complex conjugate. The planes of the
structure $\Pi$ are named the {\em principal planes} of a type D
Weyl tensor. On the other hand, a type D Weyl tensor can be
characterized as follows \cite{fms}:

\begin{lemma}
A space-time is of Petrov-Bel type D if, and only if, the self-dual
Weyl tensor satisfies:
\begin{equation}  \label{type-D}
a \not= 0 \, ,  \qquad \quad {\cal W}^2 - \frac{b}{a}\, {\cal W} -
\frac{a}{3}\, {\cal G} = 0
\end{equation}
where $a \equiv  \Tr {\cal W}^2$ and $b \equiv \Tr {\cal W}^3$.
\end{lemma}

In a type D aligned Einstein-Maxwell solution the energy tensor $T$
is of the electromagnetic type and its principal planes are those of
the Weyl tensor. In this case the energy and Ricci tensors differ in
a constant and, consequently, the Ricci tensor $R$ must satisfy the
Rainich algebraic conditions (\ref{Rainich-2-a}).

In order to impose the alignment requirement, the following two
conditions must hold: (i) $R = - \kappa \Pi$, $\Pi$ being the
principal structure of the Weyl tensor, and (ii) The self dual Weyl
tensor takes the form (\ref{type-D-canonica}), ${\cal U}$ being the
(self-dual) canonical unitary 2-form of the electromagnetic field.

If we consider a frame of normalized SD bivectors $\{{\cal U}_i\}$,
${\cal U}_1 = {\cal U}$, we have that the symmetric tensors $\{{\cal
U}_i \cdot \bar{\cal U}_j\}$ define a (complex) basis of the space
of the traceless symmetric tensors, and then the Ricci tensor may be
written as $R = \sum R^{ij}{\cal U}_i \cdot \bar{\cal U}_j$.
Moreover, any self-dual double 2-form $P$ can be spanned as $P =
\sum P^{ij} {\cal U}_i \otimes {\cal U}_j$. Then, some
straightforward algebraic calculations enable the condition for the
Ricci and Weyl tensors to have a given associated structure to be
written as follows:

\begin{lemma} \label{lemma-U}
(i) A space-time is algebraically of Einstein-Maxwell type with
unitary principal bivector ${\cal U}$ if, and only if, the Ricci
tensor $R$ satisfies:
\begin{equation}
R \not= 0 \, ,  \qquad \tr R = 0 \, , \qquad R \cdot {\cal U} =
{\cal U} \cdot R
\end{equation}

(ii) A space-time is of Petrov-Bel type D with principal structure
$\Pi$ if, and only if, the self-dual Weyl tensor ${\cal W}$
satisfies:
\begin{equation} \label{aligned-sec2-0}
a \not=0 \, , \qquad \Pi_{(\alpha}^{\mu} {\cal P}_{\beta) \mu \gamma
\delta} = 0 \, , \qquad {\cal P} \equiv {\cal W} + \frac{b}{a}\,
{\cal G}
\end{equation}
\end{lemma}

From this lemma we easily obtain the alignment condition in terms of
the Ricci and Weyl tensors:

\begin{lemma}
A Ricci tensor $R$ of non-null electromagnetic type and a Petrov-Bel
type D Weyl tensor $W$ have aligned principal planes if, and, only
if, they satisfy:
\begin{equation} \label{aligned-sec2}
a \not=0 \, , \qquad  R_{(\alpha}^{\mu} {\cal P}_{\beta) \mu \gamma
\delta} = 0 \, , \qquad {\cal P} \equiv {\cal W} + \frac{b}{a}\,
{\cal G}
\end{equation}
where ${\cal W} = \frac12(W - \ci *W)$ is the self-dual Weyl tensor
and $a \equiv  \Tr {\cal W}^2$, $b \equiv \Tr {\cal W}^3$.
\end{lemma}

Until now we have shown that the algebraic restraints for a type D
aligned Einstein-Maxwell solution are given by the Rainich
conditions (\ref{Rainich-2-a}) on the Ricci tensor, the equations
(\ref{type-D}) characterizing a type D Weyl tensor and the alignment
restriction (\ref{aligned-sec2}). Nevertheless, all these conditions
are excessive. Indeed, from lemma \ref{lemma-U},  a simple
calculation shows that, under the alignment constraint
(\ref{aligned-sec2}), the second condition in (\ref{Rainich-2-a}) on
the Ricci tensor is equivalent to restrictions (\ref{type-D}) on the
Weyl tensor. Thus we can state:

\begin{proposition} \label{prop-sec2}
The necessary and sufficient conditions for $g$ to be a type D
metric with principal planes aligned with a source tensor of
electromagnetic type is that the Ricci tensor $R = R(g)$ and the
self-dual Weyl tensor ${\cal W} = {\cal W}(g)$ satisfy the algebraic
restrictions {\em (\ref{Rainich-2-a})} and {\em
(\ref{aligned-sec2})}.
\end{proposition}

Evidently, in the above proposition we can substitute the second
condition in (\ref{Rainich-2-a}) by condition (\ref{type-D}) that
characterizes a type D Weyl tensor.

\section{Differential conditions}
\label{sec-differential-conditions}

Once we have obtained the conditions that algebraically characterize
the type D aligned Einstein-Maxwell solutions we will go on to study
the differential constraints in this section. As pointed out in the
introduction the conservation of the energy tensor, $\nabla \cdot T
= 0$, is a general consequence of the field equations. Consequently,
a priory we must only impose the differential Rainich equation
(\ref{Rainich-2-bc}). Nevertheless, we will show now that this
condition is unnecessary, that is, it is also a consequence of the
field equations.

We begin this analysis bearing in mind the {\it Maxwell-Rainich
equations}. Let $\Phi$ and $\Psi$ be the {\em expansion vector} and
the {\em rotation vector} of the principal planes of an
electromagnetic field, defined as \cite{fsY,fsU}:
\begin{equation} \label{Phi-Psi}
\begin{array}{l}
\Phi  \equiv \Phi[U] \equiv *U(\delta*U) - U(\delta U)  \\[1mm]
\Psi \equiv \Psi[U] \equiv *U(\delta U) + U(\delta *U)
\end{array}
\end{equation}
Then, we have \cite{rai,cff}:
\begin{lemma}  \label{lemma-maxrai}
In terms of the {\em intrinsic elements} $(U,\phi,\psi)$ of a
non-null Maxwell field, the source-free Maxwell equations, $\delta F
=0$, $\delta *F =0$, take the expression:
\begin{equation} \label{maxwell-rainich}
\dif \phi = \Phi[U] \, , \qquad \qquad   \dif \psi = \Psi[U]
\end{equation}
\end{lemma}

When $F$ is solution of the source-free Maxwell equations, one says
that $U$ defines a {\it Maxwellian structure}. Besides, when the
Maxwell-Minkowski energy tensor $T$ associated with a non-null
2--form is divergence--free, the underlying 2+2 structure is said to
be {\em pre-Maxwellian} \cite{deb}. The conservation of $T$ is
equivalent to the first of the Maxwell-Rainich equations
(\ref{maxwell-rainich}) \cite{fsY}. Then, from these equations we
obtain the following result \cite{rai,fsY}:
\begin{lemma}
(i) A {\em 2+2} structure is Maxwellian if, and only if, the
expansion and the rotation are closed 1--forms, namely the canonical
2--form $U$ satisfies:
\begin{equation}\label{max2}
\mbox{\rm d} \Phi[U]  = 0 \, , \qquad  \qquad \mbox{\rm d} \Psi[U] =
0
\end{equation}

(ii) A {\em 2+2} structure is pre-Maxwellian if, and only if, the
canonical 2--form $U$ satisfies the first equation in {\rm
(\ref{max2})}.
\end{lemma}

This lemma enables us to obtain the differential constraints for the
generic Rainich theory. Indeed, the algebraic conditions guarantee
that a family of 2--forms $F$ can be associated with the Ricci
tensor (see lemma \ref{lemma-FT}), and one of them must verify the
Maxwell-Rainich equations (\ref{maxwell-rainich}). The first one is
the conservation requirement, that is,  a consequence of the field
equation. The second one establishes that $\Psi$ is a closed
1--form, condition that gives the (differential) equation
(\ref{Rainich-2-bc}) if we write the rotation vector $\Psi$ in terms
of the Ricci tensor \cite{fsY}.

In showing that this differential condition is unnecessary under the
alignment requirement an important property of the type D aligned
Einstein-Maxwell solution plays a central role: the two double
Debever principal directions determine two shear-free geodesic null
congruences (umbilical principal planes). Indeed, the type D aligned
Einstein-Maxwell space-times with a non umbilical principal
structure were studied by Pleba\'nski and Hacyan \cite{ple-hac}.
They looked for solutions with cosmological constant and they found
the 'exceptional' metrics, which are solutions with a non-vanishing
cosmological constant. Thus, {\em a type D aligned Einstein-Maxwell
solution without cosmological constant has shear-free and geodesic
null principal directions}. We will show now that the umbilical
condition is equivalent to a pair of algebraic constraints.

Let us consider an Einstein-Maxwell Ricci tensor, $R = - \kappa
\Pi$, aligned with the type D Weyl tensor (\ref{type-D-canonica}),
$\Pi = 2 {\cal U} \cdot \bar{\cal U}$. Then, the Bianchi equations
take the expression:

\begin{eqnarray}
[(3 \alpha - \kappa)^2+9 \beta^2] \tau_{v} = 0 \, , \qquad [(3
\alpha + \kappa)^2+9 \beta^2] \tau_{h} = 0 \label{bianchi-umb}
\\[2mm]
\dif \kappa = 2 \kappa \Phi \label{bianchi-con} \\[2mm]
2 \dif \alpha = 3(\alpha \Phi - \beta \Psi) + \kappa \Pi(\Phi) \, ,
\qquad 2 \dif \beta = 3(\beta \Phi + \alpha \Psi) + \kappa \Pi(\Psi)
\label{bianchi-3}
\end{eqnarray}
where $\alpha$ and $\beta$ are the real and imaginary parts of the
double Weyl eigenvalue, $\rho = \alpha + \ci \beta$, $\Phi$ and
$\Psi$ are the expansion and rotation vectors given in
(\ref{Phi-Psi}), and $\tau_v$ and $\tau_h$ are the traceless part of
the symmetric second fundamental form of the principal planes (see,
for example \cite{fsU} for a specific definition). Considering $2
\kappa = e^{2 \phi}$, equation (\ref{bianchi-con}) becomes the first
of the Maxwell-Rainich equations (\ref{maxwell-rainich}) and,
consequently, it is precisely the conservation equation.

When $\tau_v$ and $\tau_h$ vanish the principal planes define an
umbilical structure \cite{fsU}, and this geometric property is
equivalent to the shear-free and geodesic character of the two null
principal vectors \cite{fsU,fsD}. Thus, from the equations
(\ref{bianchi-umb}) we have that, if in a type D aligned
Einstein-Maxwell space-time the Ricci eigenvalue $\kappa$ and the
Weyl eigenvalue $\rho= \alpha + \ci \beta$ satisfy
\begin{equation}  \label{alg-umb}
(3 \alpha - \kappa)^2+9 \beta^2 \not= 0 \, , \qquad (3 \alpha +
\kappa)^2+9 \beta^2 \not= 0 \, ,
\end{equation}
then the principal planes are umbilical, that is, the Debever
principal directions define shear-free geodesic null congruences.

Now we show that the converse statement also holds: the solutions
with umbilical principal planes satisfy, necessarily, the algebraic
restrictions (\ref{alg-umb}). Indeed, if the first (resp., second)
condition in (\ref{alg-umb}) does not hold, one has $\beta = 0$ and
$3 \alpha = \kappa$ (resp., $3 \alpha = - \kappa$). Then, Bianchi
identities (\ref{bianchi-con}) and (\ref{bianchi-3}) imply that
$\Phi = 0$ and $v(\Psi) = 0$ (resp., $\Phi = 0$ and $h(\Psi) = 0$).
In \cite{fsU} we have obtained the expression of the Ricci tensor in
a space-time with an umbilical structure. In this case the Ricci
tensor depends, up to two scalars, on the expansion and rotation
vectors $\Phi$ and $\Psi$. When $\Phi = 0$ the Ricci tensor becomes
\cite{fsU}:
\begin{equation}
R = r_v v + r_h h + \Psi \otimes \Psi - *U \cdot {\cal L}_{\Psi}g
\cdot U - U \cdot {\cal L}_{\Psi}g \cdot *U
\end{equation}

If one now imposes that $R = - \kappa \Pi = - \kappa(v - h)$, one
obtains that $\Psi=0$. Thus, the principal planes are umbilical,
minimal ($\Phi = 0$) and integrable ($\Psi=0$) and, consequently,
$g$ becomes a product metric \cite{fsU}. Now $R = r_v v + r_h h$ and
the traceless condition implies $r_v + r_h=0$. But, for a product
metric, this condition leads to a vanishing Weyl tensor, that
contradicts our type D hypothesi. Consequently, under the umbilical
property, (\ref{alg-umb}) applies and we can state:

\begin{lemma} \label{lemma-umb-is-alg}
In a type D aligned Einstein-Maxwell space-time the Debever
principal directions define shear-free geodesic null congruences if,
and only if, the Ricci eigenvalue $\kappa$ and the Weyl eigenvalue
$\rho= \alpha + \ci \beta$ satisfy {\rm (\ref{alg-umb})}.
\end{lemma}

Elsewhere \cite{fsU} we have studied the integrability conditions
for the umbilical condition and we have obtained interesting
consequences for the different Petrov-Bel types. A detailed analysis
of the type D space times leads to the following (see proposition 10
in reference \cite{fsU}):

\begin{lemma} \label{pro-max-iff}
In a type D space-time whose Debever null principal directions are
shear-free and geodesic (the principal planes are umbilical), if the
principal structure is pre-Maxwellian, $\dif \Phi = 0$, then it is
Maxwellian, $\dif \Phi = \dif \Psi = 0$.
\end{lemma}

This lemma implies that, under the umbilical condition, the
differential Rainich equation (\ref{Rainich-2-bc}) is a consequence
of the field equations. But lemma \ref{lemma-umb-is-alg} states that
the umbilical (differential) condition can be expressed,
equivalently, by means of the (algebraic) constraint
(\ref{alg-umb}). On the other hand, remembering that, as a
consequence of the work by Pleba\'nski and Hacyan \cite{ple-hac},
all the type D aligned Einstein-Maxwell solutions without
cosmological constant have umbilical principal planes, and taking
into account proposition \ref{prop-sec2}, we can state:

\begin{proposition} \label{prop-sec3}
The necessary and sufficient conditions for $g$ to be a type D
aligned Einstein-Maxwell solution is that the Ricci tensor $R =
R(g)$ and the self-dual Weyl tensor ${\cal W} = {\cal W}(g)$ satisfy
the algebraic restrictions {\em (\ref{Rainich-2-a})}, {\em
(\ref{aligned-sec2})} and {\em (\ref{alg-umb})}.
\end{proposition}

We can make the characterization given in the above proposition more
explicit with the following considerations. From the expression
(\ref{type-D-canonica}) we obtain the real imaginary part of the
Weyl eigenvalue as
$$
\displaystyle \alpha \equiv - \frac{AC+BD}{A^2+B^2} \, ,  \qquad
\beta \equiv \frac{AD-BC}{A^2+B^2} \, ,
$$
where $A,B,C,D$ are the real invariants of the Weyl tensor that are
related to the complex ones by $2a = A - \ci B$, $2b = C - \ci D$.

On the other hand, the alignment equation (\ref{aligned-sec2}) is
equivalent to its real part since the Ricci tensor $R$ is real and
${\cal P}$ is a self-dual double 2--form. Thus, taking the real part
in (\ref{aligned-sec2}) we can write this condition in terms of the
(real) Weyl tensor and its real invariants. Finally, the two
constraints (\ref{alg-umb}) state, equivalently, that $6 a \not= \tr
R^2$. Taking into account all these considerations proposition
\ref{prop-sec3} implies our main theorem \ref{teo-our-1} settled in
the introduction.

Theorem \ref{teo-our-1} characterizes the family of type D aligned
Einstein-Maxwell space-times by means of {\it algebraic} conditions.
In order to complete the Eins\-tein-Maxwell solution we need to
obtain the Maxwell field $F$ associated with the Maxwell-Minkowski
energy tensor $T$. If we take into account the Maxwell-Rainich
equations (\ref{maxwell-rainich}), the expression $\Psi(R)$ in
(\ref{Rainich-2-bc}) of the rotation vector in terms of the Ricci
tensor, and the second statement in lemma \ref{lemma-FT}, we arrive
to the following result:

\begin{proposition}
Let $g$ be a metric tensor satisfying the conditions {\rm
(\ref{Rainich-3-a})} and {\rm (\ref{aligned-3})} of theorem {\rm
\ref{teo-our-1}}. Then, a function $\psi$ exists such that $\dif
\psi = \Psi(R)$ where $\Psi(R)$ is given in {\rm
({\ref{Rainich-2-bc})}. Let us consider the 2--form $F = \cos \psi
F^{\circ} + \sin \psi
*F^{\circ}$, where $F^{\circ} = F^{\circ}(R)$ is given in {\rm (\ref{F0}})}.
Then the pair $(g,F)$ is an Einstein-Maxwell solution.
\end{proposition}

\section{On the solutions with cosmological constant and on the vacuum
limit}  \label{sec-ending}

The explicit integration of the Einstein-Maxwell equations for the
family of metrics characterized in theorem \ref{teo-our-1} has been
obtained by several authors (see \cite{ple-dem,weir-kerr,deb-kam-mc}
and references therein). This family of solutions, also including
their counterpart with cosmological constant, has been named the
${\cal D}$-metrics \cite{deb-kam-mc}. They can be deduced from the
Pleba\'nsky and Demia\'nski \cite{ple-dem} line element by means of
several limiting procedures (see \cite{kra} and references therein
and \cite{griff-pod} for a recent analysis).

The Rainich-like characterization of the ${\cal D}$-metrics easily
follows from the analysis presented in the previous sections.
Indeed, the algebraic conditions applies if we replace the Ricci
tensor by its traceless part and impose a constant trace. On the
other hand, in showing that the differential Rainich condition is
unnecessary we have used the Bianchi equations
(\ref{bianchi-umb},\ref{bianchi-con},\ref{bianchi-3}) that take
exactly the same expression when the cosmological constant is
present. Thus, we can state:

\begin{theorem}  \label{teo-our-2}
A metric tensor $g$ is a charged ${\cal D}$-metric if, and only if,
its Ricci and Weyl tensors,  $R \equiv R(g)$ and $W = W(g)$ satisfy
the conditions:
\begin{equation} \label{Rainich-3-a-tilde}
\dif \tr R = 0, \qquad \quad   4 \tilde{R}^2 = \tr \tilde{R}^2\, g
\not= 0, \qquad \quad \tilde{R}(x,x) > 0   \label{alg}
\end{equation}
\begin{equation} \label{aligned-3-tilde}
R_{(\alpha}^{\mu} P_{\beta) \mu \gamma \delta} = 0 \, , \qquad A^2 +
B^2 \not= 0 \, , \qquad B^2 + ( 3 A - \tr \tilde{R}^2)^2 \not= 0,
\end{equation}
where $P$, $A$, and $B$ depend on the Weyl tensor as given in
theorem {\rm \ref{teo-our-1}}, $\tilde{R}$ is the traceless part of
the Ricci tensor, $\tilde{R} \equiv R - \frac14 \tr R \, g$, and
where $x$ is an arbitrary time-like vector.
\end{theorem}

It is worth remarking that theorem \ref{teo-our-2} does not give the
Rainich-like characterization of the whole set of type D aligned
Einstein-Maxwell solutions with cosmological constant. Indeed,
theorem \ref{teo-our-2} characterizes the ${\cal D}$-metrics, and
this family does not contain the exceptional metrics by Pleba\'nski
and Hacyan \cite{ple-hac}.

Let us also note that the space-times which theorems \ref{teo-our-1}
and \ref{teo-our-2} characterize do not include the vacuum limit.
Indeed, the algebraic Rainich conditions (\ref{Rainich-3-a-tilde})
avoid the vacuum case $R = 0$ and, moreover, the alignment condition
(first in (\ref{aligned-3-tilde})) holds identically and does not
guarantee a Weyl tensor of type D.

In order to obtain a Rainich-like characterization of the ${\cal
D}$-metrics that also includes the vacuum solutions we must replace
the second condition in (\ref{Rainich-3-a-tilde}) by the conditions
(\ref{type-D}) for a type D Weyl tensor. Then, we obtain an
alternative characterization that we give in complex vectorial
formalism:

\begin{theorem}  \label{teo-our-3}
A metric tensor $g$ is a ${\cal D}$-metric (including the vacuum
limit) if, and only if, its Ricci and Weyl tensors,  $R \equiv R(g)$
and $W = W(g)$ satisfy the conditions:
\begin{equation}
\dif \tr R = 0, \qquad \qquad  \tilde{R}(x,x) \geq 0
\label{alg-our-3}
\end{equation}
\begin{equation}  \label{type-D-our-3}
a \not= 0 \, ,  \qquad \quad {\cal W}^2 - \frac{b}{a}\, {\cal W} -
\frac{a}{3}\, {\cal G} = 0
\end{equation}
\begin{equation} \label{aligned-sec2-our-3}
R_{(\alpha}^{\mu} {\cal P}_{\beta) \mu \gamma \delta} = 0 \, ,
\qquad  6 a \not= \tr \tilde{R}^2 \, ,
\end{equation}
where
$$
\begin{array}{ll}
\tilde{R} \equiv R - \frac14 \tr R \, g \, , & \qquad \displaystyle
{\cal P} \equiv {\cal W} + \frac{b}{a}\, {\cal G} \, , \qquad  a
\equiv \Tr {\cal W}^2 \, , \quad b \equiv \Tr {\cal W}^3 \, ,
\\[3mm]
{\cal G} \equiv \frac12(G - \ci \eta) \, , & \qquad  {\cal W} \equiv
\frac12(W - \ci *W) \, ,
\end{array}
$$
and where $G = \frac12 g \wedge g$ is the metric on the {\rm
2}--forms space, $\eta$ is the metric volume element and $x$ is an
arbitrary time-like vector.
\end{theorem}

The main results of this work (theorems \ref{teo-our-1},
\ref{teo-our-2} and \ref{teo-our-3}) offer an {\it intrinsic}
(depending solely on the metric tensor) and {\it explicit}
characterization of an important family of known solutions of the
Einstein equations: the ${\cal D}$-metrics. This kind of
characterization has already been given for two subsets of this
family: the static type D vacuum solutions \cite{fsS} and their
charged counterpart \cite{fsD}. The results in these quoted works
have allowed us to obtain an intrinsic and explicit labeling of
every metric of these families and, in particular, to characterize
the Schwarzschild and Reissner-Nordstr\"om black holes \cite{fsS}
\cite{fsD}.

In a similar way, the characterization of the ${\cal D}$-metrics
that we give in the present work is a first step in labeling every
particular solution in this family. This task and, in particular,
the intrinsic characterization of the Kerr and Kerr-Newman black
holes will be tackled elsewhere \cite{fsEM-JMP}.

%
{\bf Acknowledgements} This work has been partially supported by the
Spanish Ministerio de Educaci\'on y Ciencia, MEC-FEDER project
FIS2006-06062.

\appendix

\section{Notation}

\begin{enumerate}
\item
{\bf Products and other formulas involving 2-tensors $A$ and $B$}:
\begin{enumerate}
\item
Composition as endomorphisms: $A \cdot B$,
\begin{equation}
(A \cdot B)^{\alpha}_{\ \beta} = A^{\alpha}_{\ \mu} B^{\mu}_{\
\beta}
\end{equation}
\item
Square and trace as an endomorphism:
\begin{equation}
A^2 = A \cdot A \, , \qquad \tr A = A^{\alpha}_{\ \alpha}.
\end{equation}
\item
Action on a vector $x$, as an endomorphism $A(x)$, and as a
quadratic form $ A(x,x)$:
\begin{equation}
A(x)^{\alpha} = A^{\alpha}_{\ \beta} x^{\beta}\, , \qquad A(x,x) =
A_{\alpha \beta} x^{\alpha} x^{\beta} \, .
\end{equation}
\item
Exterior product as double 1-forms:  $(A \wedge B)$,
\begin{equation}
(A \wedge B)_{\alpha \beta \mu \nu} = A_{\alpha \mu} B_{\beta \nu} +
A_{\beta \nu} B_{\alpha \mu} - A_{\alpha \nu} B_{\beta \mu} -
A_{\beta \mu} B_{\alpha \nu} \, .
\end{equation}
\end{enumerate}
\item
{\bf Products and other formulas involving double 2-forms $P$ and
$Q$}:
\begin{enumerate}
\item
Composition as endomorphisms of the bivectors space: $P \circ Q$,
\begin{equation}
(P \circ Q)^{\alpha \beta}_{\ \ \rho \sigma} = \frac12 P^{\alpha
\beta}_{\ \ \mu \nu} Q^{\mu \nu}_{\ \ \rho \sigma}
\end{equation}
\item
Square and trace as an endomorphism:
\begin{equation}
P^2 = P \circ P \, , \qquad \Tr P = \frac12 P^{\alpha \beta}_{\ \
\alpha \beta}.
\end{equation}
\item
Action on a bivector $X$, as an endomorphism $P(X)$, and as a
quadratic form $P(X,X)$,
\begin{equation}
P(X)^{\alpha \beta} = \frac12 P^{\alpha \beta}_{\ \ \mu \nu} X^{\mu
\nu} \, , \qquad P(X,X) = \frac14 P_{\alpha \beta \mu \nu} X^{\alpha
\beta} X^{\mu \nu}.
\end{equation}
\end{enumerate}
\item
{\bf Other products and operators}:
\begin{enumerate}
\item
The Hodge dual operator is defined as the action of the metric
volume element $\eta$ on a bivector $F$ and a double 2-form $W$:
\begin{equation}
*F = \eta(F)  \, , \qquad  *W = \eta \circ W \, .
\end{equation}
\item
For a tensor $t$, $*(t)$ denotes the action of the Hodge dual
operator on the skew-symmetric part of $t$. For example,
\begin{equation}
*(\nabla T \cdot T)_{\alpha} = \frac16 \eta_{\alpha}^{\ \beta \mu \nu}
\nabla_{\beta} T_{\mu \lambda}\, T^{\lambda}_{\ \nu} \, .
\end{equation}
Note that the symbol $\cdot$ indicates, as in the endomorphism
composition, the contraction of the adjacent indices of the
tensorial product.
\end{enumerate}
\end{enumerate}

\end{document}